# What Work is AI Actually Doing? Uncovering the Drivers of Generative AI Adoption

Peeyush Agarwal[a*], Harsh Agarwal[b*], and Akshat Rana[a]

[a]Netaji Subhas University of Technology (NSUT), Delhi, India

[b]Adobe Inc., Noida, India

[a*]peeyush.agarwal.ug22@nsut.ac.in; [b*]harshaga@adobe.com

## Abstract

**Purpose**

The rapid integration of artificial intelligence (AI) systems like ChatGPT, Claude AI, etc., has a deep impact on how work is done. Predicting how AI will reshape work requires understanding not just its capabilities, but how it is actually being adopted. This study investigates which intrinsic task characteristics drive users' decisions to delegate work to AI systems.

**Methodology**

This study utilizes the Anthropic Economic Index dataset of four million Claude AI interactions mapped to O*NET tasks. We systematically scored each task across seven key dimensions: Routine, Cognitive, Social Intelligence, Creativity, Domain Knowledge, Complexity, and Decision Making using 35 parameters. We then employed multivariate techniques to identify latent task archetypes and analyzed their relationship with AI usage.

**Findings**

Tasks requiring high creativity, complexity, and cognitive demand, but low routineness, attracted the most AI engagement. Furthermore, we identified three task archetypes: Dynamic Problem Solving, Procedural & Analytical Work, and Standardized Operational Tasks, demonstrating that AI applicability is best predicted by a combination of task characteristics, over individual factors. Our analysis revealed highly concentrated AI usage patterns, with just 5% of tasks accounting for 59% of all interactions.

**Originality**

This research provides the first systematic evidence linking real-world generative AI usage to a comprehensive, multi-dimensional framework of intrinsic task characteristics. It introduces a data-driven classification of work archetypes that offers a new framework for analyzing the emerging human-AI division of labor.

# 1. Introduction

The recent proliferation of generative Artificial Intelligence (AI) systems marks a pivotal moment in the history of technology and work. Unlike previous technological waves that were followed by decades of research to understand their impact, we now have the rare opportunity to observe AI's integration into the economy in real-time (Eloundou et al., 2024). The trajectory of human progress has been shaped by our capacity to create tools that extend our capabilities, from stone implements to steam engines, each redefining human potential (Bresnahan and Trajtenberg, 1995). In this continuum, AI represents a transformative milestone, with Large Language Models (LLMs) like ChatGPT, Claude AI, and Gemini promising to fundamentally reshape economies and labor markets by augmenting human cognitive capabilities on an unprecedented scale (Felten et al., 2023). These systems exhibit general-purpose abilities that extend far beyond narrow functions, from natural language processing and code generation to complex reasoning and creative synthesis (Eloundou et al., 2024). Within months of their release, millions of people began using these tools for writing, coding, analysis, and creative tasks (Bick et al., 2024). This represents a fundamental shift in how we think about AI's role in the economy.

However, understanding the impact of AI on the labor market requires moving beyond broad predictions of occupational disruption. Early analyses often focused on entire occupations or industries, asking whether jobs like "accountant" or "writer" would be automated. But occupations are not monolithic entities. They are collections of diverse tasks, each with different characteristics. For example, a market research analyst performs routine data collection, complex statistical analysis, creative report writing, and interpersonal client presentations. AI is unlikely to affect each of these activities in the same way(Bonney et al., 2024). This leads to a fundamental question: what determines which specific tasks are most amenable to AI assistance? To answer this, we need a task-based approach, as pioneered by (Autor et al., 2003). By breaking down jobs into their constituent activities, we can identify which work tasks are susceptible to automation, which are ripe for human-AI collaboration, and where human expertise remains irreplaceable (Acemoglu and Restrepo, 2019).

Until recently, empirical analysis at this granular level was constrained by a lack of data on real-world generative AI usage. A significant advancement comes from Anthropic's Economic Index dataset (Handa et al., 2025).

It maps millions of user interactions with the Claude AI model to the standardized task taxonomy of the U.S. Department of Labor's Occupational Information Network (O*NET) (Peterson et al., 2001). This dataset provides the first large-scale empirical evidence of how LLMs are being used across the spectrum of occupational tasks.

While this work shows which tasks have high levels of AI usage, it doesn't explain why these patterns exist or what characteristics make certain tasks more suitable for AI assistance. Our study addresses this critical gap by developing and operationalizing a multi-dimensional framework to quantify the intrinsic properties of these tasks. We hypothesize that a task's suitability for AI interaction is a function of its underlying characteristics. Our framework assesses tasks across seven key dimensions: **Routine** (*how standardized the procedures are*), **Cognitive** (*intellectual complexity*), **Social Intelligence** (*interpersonal and emotional skills*), **Creativity** (*innovative and generative demands*), **Domain Knowledge** (*specialized expertise needed*), **Complexity** (*multifaceted problem-solving*), and **Decision Making** (*judgment and* *choice-making*). We further break down each of these dimensions into 5 parameters and then use a large language model to score the O*NET tasks across these 35 parameters.

This study makes three primary contributions to our understanding of AI's real-world impact. First, we provide the first systematic evidence linking real-world AI usage patterns to intrinsic task characteristics. Second, we introduce a comprehensive multi-dimensional framework for characterizing occupational tasks that captures the multifaceted nature of modern AI capabilities. Third, we identify distinct task archetypes that reveal deeper patterns in AI adoption than individual characteristics alone. These insights have practical implications for multiple stakeholders, including policymakers, organizations, and individuals. The study can inform evidence-based policy-making, guide strategic decisions on AI implementation, and illuminate career development pathways in an AI-augmented economy. To guide our inquiry, we pose the following research questions:

1. What is the distribution and concentration of AI usage across occupational tasks?
2. What specific task characteristics are the strongest predictors of AI adoption?
3. Can tasks be grouped into meaningful archetypes based on their characteristic profiles, and how does AI usage vary across these archetypes?

The remainder of this paper is organized as follows. Section II reviews the relevant literature on AI's economic impact and task-based analysis. Section III details our data sources and analytical methodology. Section IV presents the empirical results of our analysis. Section V discusses the interpretation and broader implications of these results. Finally, Section VI concludes with a summary of our contributions and directions for future research.

## 2. Literature Review

This section reviews the theoretical and empirical foundations underlying our analysis of AI adoption patterns across work tasks. We examine the evolution of research on AI's labor market impact, the emergence of task-based analytical frameworks, challenges in measuring AI adoption, and the conceptual basis for our task characteristics framework. Our review reveals both the progress made in understanding AI's economic impact and the critical gaps that remain in our knowledge.

*2.1. AI and the Future of Work*

The debate over AI's impact has shifted from forecasts of mass job displacement (Frey and Osborne, 2017) to a view of AI as an augmentation tool that enhances productivity and reshapes professional roles (Brynjolfsson et al., 2025; Sarala et al., 2025). This reframing is challenged by a critical perspective advocating for the use of AI not just to preserve jobs, but to lighten labor and reduce work hours through a broader democratization of the technology (Spencer, 2025). Furthering this forward-looking debate, some economic models suggest that sufficiently advanced AI could become "transformative" by fundamentally altering long-run growth dynamics and potentially breaking historical trends like a stable labor share, raising large-scale questions about income distribution in a future of recursively self-improving capital (Trammell and Korinek, 2023).

*2.2. Task-Based Approaches to Labor Market Analysis*

Earlier frameworks, notably the Skill-Biased Technical Change (SBTC) hypothesis, proved insufficient for explaining complex labor market trends, such as wage polarization (Card and DiNardo, 2002). This prompted a critical paradigm shift towards task-based analysis. The foundational insight, pioneered by Autor, Levy, and Murnane (Autor et al., 2003), is that technology does not automate whole jobs, but rather substitutes for specific tasks.

This perspective was later refined to show how automation amplifies the comparative advantage of workers in supplying non-routine problem-solving, adaptability, and creativity, while noting that this interplay can lead to a polarization of the labor market (Autor, 2015). This task-based approach, often operationalized using the O*NET database (Handel, 2016; Deming, 2017; Webb, 2020), has become the dominant framework. Modern extensions of this model, particularly from Acemoglu and Restrepo (Acemoglu and Restrepo, 2019), frame technology as a dual force that both displaces labor from existing tasks and reinstates it in new ones. This dynamic adaptation is better understood through theories of technology appropriation, which posit that users and organizations mutually shape how AI is integrated into workflows (Corvello, 2025).

*2.3. Characterizing Task Demands*

As AI capabilities advance, the traditional binary of routine versus non-routine tasks (Deranty, 2024) becomes insufficient. While the automation of routine work remains a central theme (Barenkamp et al., 2020; Upreti et al., 2024), the frontier of AI research has expanded dramatically to domains long considered uniquely human. This includes complex cognitive abilities such as pattern recognition (Tolan et al., 2021; Tyson et al., 2022); creativity, where workers face a tension between AI as a co-creation tool and a market-driven necessity (Wingstrom et al., 2024; Idowu et al., 2024; Öztaş and Arda, 2025); and even social intelligence (Mathur et al., 2024). The integration of GenAI can psychologically threaten workers' sense of competence and professional identity by mimicking these skills (Hermann et al., 2025). At the same time, AI's application to high-stakes problems highlights the importance of domain knowledge (Miller et al., 2024; Johnson et al., 2022), managing task complexity, and augmenting human decision making (Soori et al., 2024; Macnamara et al., 2024). Therefore, understanding modern AI's impact requires moving to a multi-dimensional framework capable of capturing the full character of contemporary work.

*2.4. Measuring AI Adoption and Impact*

A persistent challenge in AI research has been the difficulty in measuring its real-world adoption. Because AI is a general-purpose technology, most studies rely on indirect measures like patent data (Pairolero et al., 2025), firm surveys (Singla et al., 2023; Company, 2024), or the analysis of AI skills in job postings (Green et al., 2025; Hampole et al., 2025). These methods track innovation, investment, and skill demand but do not capture direct application to tasks. A different, user-level approach uses frameworks like the Technology Acceptance Model (TAM) to show that individual adoption is driven by perceived utility and moderated by user demographics (Ma et al., 2025). This micro-level perspective on user behavior complements the macro-level economic proxies, but a gap remains in linking either approach to the intrinsic nature of the work tasks themselves.

*2.5. Identified Gaps*

While the literature has established the primacy of task-based analysis (Autor et al., 2003; Acemoglu and Autor, 2011; Acemoglu and Restrepo, 2019) and the need for multi-dimensional task characterizations (Tolan et al., 2021; Wingstrom et al., 2024), our review reveals three critical gaps that this study directly addresses:

1. **Lack of Empirical Usage Data**: Current research largely measures potential AI exposure through proxies such as expert ratings (Felten et al., 2021) or market proxies such as startup activity (Fenoaltea et al., 2024) to establish potential AI exposure; they do not capture the ground truth of adoption.

2. **Need for Multi-faceted Analysis**: Emerging data on real-world AI use has not yet been systematically correlated with a comprehensive, multi-dimensional profile of task characteristics. This prevents an analysis of which intrinsic properties of a task attract AI augmentation.

3. **Moving from 'What' to 'Why':** Consequently, the key gap remains in moving from identifying what tasks AI is used for to quantitatively exploring why. Understanding the relationship between the frequency of AI interaction and the underlying nature of a task is essential for providing an explanatory model for current adoption patterns.

## 3. Materials and Methods

This section describes our approach to investigating the relationship between real-world AI usage and intrinsic task characteristics. Our methodology integrates a large-scale AI usage dataset with a novel, multidimensional task characteristics framework, which is then analyzed using multiple descriptive, multivariate, and clustering techniques.

*3.1. Data Sources*

Our analysis is built upon the integration of two primary data sources: the O*NET Database and the Anthropic Economic Index Dataset.

The **Occupational Information Network (O*NET) Database** provides the foundational taxonomy of occupational tasks for our analysis. O*NET breaks down hundreds of occupations into thousands of standardized task descriptions such as "analyze data to determine feasibility of product proposals.".

We use the public dataset released as part of **Anthropic's Economic Index** initiative to derive our measure of AI adoption. This dataset was generated by Anthropic by analyzing millions of anonymized user conversations with the Claude.ai model. Anthropic linked these user interactions to the most relevant occupational task category as defined by the U.S. Department of Labor's O*NET database using Clio (Tamkin et al., 2024). Our key measure, **AI Usage**, represents what percentage of all AI conversations were related to each specific task. This provides a proxy for the frequency of real-world AI engagement on a task-by-task basis.

*3.2. Task Characteristics Framework*

We developed a novel, multi-dimensional framework to quantify the intrinsic characteristics of each task, aiming to understand why certain tasks attract more AI usage. Our framework builds upon established research in labor economics and cognitive science, adapting these concepts specifically to understand AI adoption patterns.

We identified seven core characteristics established in labor economics and cognitive science as critical for differentiating work: Routine, Cognitive, Social Intelligence, Creativity, Domain Knowledge, Complexity, and Decision Making. These are defined as follows:

1. **Routine:** Measures the degree to which a task is repetitive and can be completed by following a well-defined set of rules or procedures. This builds on the work of (Autor et al., 2003).
2. **Cognitive:** Assesses the level of mental effort required, including analysis, reasoning, and problem-solving. (Sweller et al., 1998) developed cognitive load theory to understand how task complexity affects learning and performance.
3. **Social Intelligence:** Gauges the extent to which a task requires interpersonal interaction, such as persuasion, empathy, and understanding social cues. (Deming, 2017) showed that jobs requiring social skills have grown substantially, and recent work by (Weinberg, 2022) suggests that social skills may be particularly durable in the AI era.
4. **Creativity:** Evaluates the need for originality, imagination, and the generation of novel ideas. While (Brynjolfsson et al., 2019) argues that AI can augment creative work, (Bessen, 2019) cautions that true creativity may remain uniquely human.
5. **Domain Knowledge:** Measures the requirement for specialized and in-depth knowledge of a particular field (e.g., medicine, law, programming) to perform the task effectively.
6. **Complexity:** Assesses the number of interdependent variables, the intricacy of the information involved, and the multifaceted nature of the task.
7. **Decision Making:** Gauges the significance and consequence of the judgments that must be made during the task.

To ensure a rigorous and transparent scoring process, we decomposed each of these seven characteristics into five specific measurable parameters. This decomposition results in a total of 35 granular parameters that form the basis of our scoring protocol, as detailed in Table 1.

| Characteristic | Parameter | Description |
| --- | --- | --- |
| Routine | Frequency of repetition | How often the task is performed in the same manner |
| | Adherence to procedures | The extent to which the task requires following established rules |
| | Predictability of outcomes | Whether the results are consistent and expected |
| | Need for supervision | The level of oversight required |
| | Susceptibility to automation | The likelihood that the task can be automated |
| Cognitive | Information processing | The amount of data or information that needs to be handled |
| | Memory requirements | The necessity to recall information or procedures |
| | Analytical thinking | The need to analyze situations or data |
| | Problem-solving | The requirement to find solutions to challenges |
| | Learning and adaptation | The need to acquire new knowledge or adjust to changes |
| Social Intelligence | Frequency of interaction | How often the task involves communicating with others |
| | Complexity of communication | The intricacy of communication required (simple instructions vs negotiations) |
| | Emotional demands | The need to manage emotions or understand others' feelings |
| | Collaboration requirements | Whether the task involves working with others |
| | Social perceptiveness | The ability to read social cues and understand social dynamics |
| Creativity | Need for innovation | The requirement to develop new ideas or methods |
| | Originality of output | The expectation for unique or novel results |
| | Flexibility in approach | The ability to use different methods or perspectives |
| | Problem-solving creativity | The need to solve problems in innovative ways |
| | Artistic or aesthetic components | Whether the task involves creating something with artistic value |
| Domain Knowledge | Specialized knowledge | The extent of specific knowledge required in a particular field |
| | Technical skills | The necessity for proficiency in certain tools or techniques |
| | Experience level | The amount of practical experience needed to perform the task |
| | Educational requirements | The level of formal education or training necessary |
| | Updating knowledge | The frequency with which one must stay current with new developments |
| Complexity | Number of components | The quantity of distinct parts or steps in the task |
| | Interrelatedness | How much the different parts of the task depend on each other |
| | Skill diversity | The variety of skills needed to complete the task |
| | Time pressure | The urgency or deadlines associated with the task |
| | Consequence of errors | The potential impact if the task is not performed correctly |
| Decision Making | Frequency of decisions | How often decisions need to be made during the task |
| | Significance of decisions | The importance or impact of the decisions |
| | Complexity of choices | The difficulty in choosing between options |
| | Information gathering | The need to collect and analyze data before deciding |
| | Autonomy in decision-making | The level of independence in making decisions |

Table 1: Task Characteristics, Parameters, and Descriptions. (Source: Authors' own work)

*3.3. LLM-Based Task Scoring*

To systematically score thousands of O*NET tasks, on all 35 parameters, we employed a large language model, Gemini 2.5 Pro (with temperature = 0). This approach aligns with the emerging "LLM-as-a-Judge" paradigm, where LLMs serve as a scalable and consistent alternative to human experts for complex evaluation tasks (Gu et al., 2024; Li et al., 2024). The methodology has been successfully applied in specialized domains that require expert-level nuance, such as legal case analysis (Ma et al., 2024). While the reliability of LLM judgments is an active area of research requiring careful methodological design (Schroeder and Wood-Doughty, 2024), the paradigm is increasingly established for scoring large and nuanced datasets. We provide the model with the official O*NET task description for each task and prompted it to assign a rating on a scale of 1-10 for each specific parameter. From these 35 granular scores, we constructed our seven primary characteristic scores. The final score for each of the seven characteristics is the arithmetic mean of its five constituent parameter scores.

*3.4. Statistical Analysis*

Our analytical strategy is designed to move from granular relationships to higher-order patterns in AI adoption. We begin by examining the direct relationship between AI Usage and our 35 granular and 7 aggregate task characteristics using Spearman's rank-order correlations and profile comparisons. To uncover the underlying structure of work, we then employ Principal Component Analysis (PCA) to reduce the dimensionality of the seven primary characteristics. Using the principal components, we perform K-Means clustering to identify empirically-derived "task archetypes". Finally, we validate the statistical distinctiveness of these archetypes and test for significant differences in AI usage across them using Multivariate Analysis of Variance (MANOVA).

## 4. Results

This section presents the empirical findings of our analysis. We first describe the overall landscape of AI usage and task characteristics. We then do a granular analysis of the specific task parameters driving adoption and then identify and analyze a robust set of task archetypes that exhibit significantly different rates of AI adoption.

*4.1. The Patterns in AI Usage and Task Characteristics*

Our analysis reveals that the current share of AI usage is highly concentrated. The distribution of AI usage across O*NET tasks is sharply right-skewed (Figure 1). Only a small number of tasks attract a significant share of AI interaction. Table 2 shows that the median task accounts for only 0.006% of AI conversations and 75% of all tasks fall below 0.017% usage. This suggests that the current adoption of AI is not yet widely distributed across all work activities.

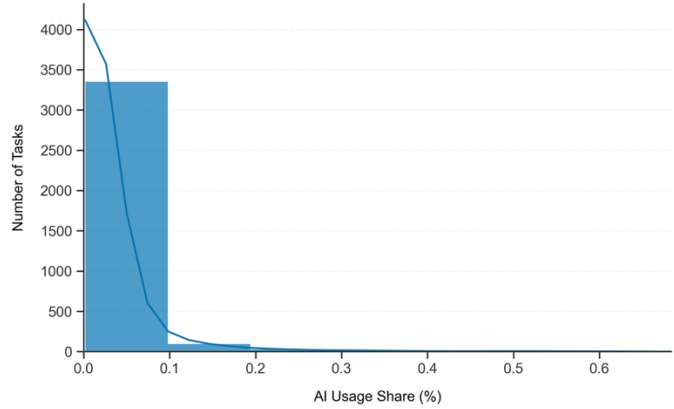

Figure 1: Distribution of AI Usage Share Across O*NET Tasks. (Source: Authors' own work)

As shown in the box plots in Figure 2a, the distributions for Cognitive (median 8.2), Complexity (median 8.0), and Decision Making (median 7.8) scores are left-skewed with high median scores. Conversely, Routine (median 4.2) is right-skewed. This profile indicates that the tasks where AI is being applied are largely representative of demanding knowledge work. Furthermore, the characteristics themselves are logically interrelated. The inter-correlation matrix (Figure 2b) reveals that Cognitive, Complexity, and Decision Making characteristics are strongly and positively correlated (all Spearman's $\rho > 0.75$), while all are strongly negatively correlated with Routine (all $\rho < -0.61$). Social Intelligence, on the other hand, exhibits weaker correlations with all characteristics. This structure provides the rationale for the multivariate analyses that follow.

|  | mean | std | min | 25% | 50% | 75% | max |
|---|---|---|---|---|---|---|---|
| **AI Usage** | 0.028 | 0.14 | 0.002 | 0.003 | 0.006 | 0.017 | 4.794 |
| **Routine** | 4.486 | 1.82 | 1 | 3 | 4.2 | 5.8 | 9.6 |
| **Cognitive** | 7.795 | 1.448 | 1 | 7.4 | 8.2 | 8.8 | 10 |
| **Social Intelligence** | 5.918 | 2.4 | 1 | 4.2 | 6.6 | 7.8 | 10 |
| **Domain Knowledge** | 7.703 | 1.27 | 1 | 7.2 | 8 | 8.6 | 10 |
| **Complexity** | 7.594 | 1.532 | 1 | 7.2 | 8 | 8.6 | 10 |
| **Creativity** | 4.994 | 2.32 | 1 | 3 | 5.4 | 6.8 | 9.8 |
| **Decision Making** | 7.444 | 1.546 | 1 | 7 | 7.8 | 8.4 | 9.8 |

Table 2: AI Usage and Task Characteristic Score Distributions. (Source: Authors' own work)

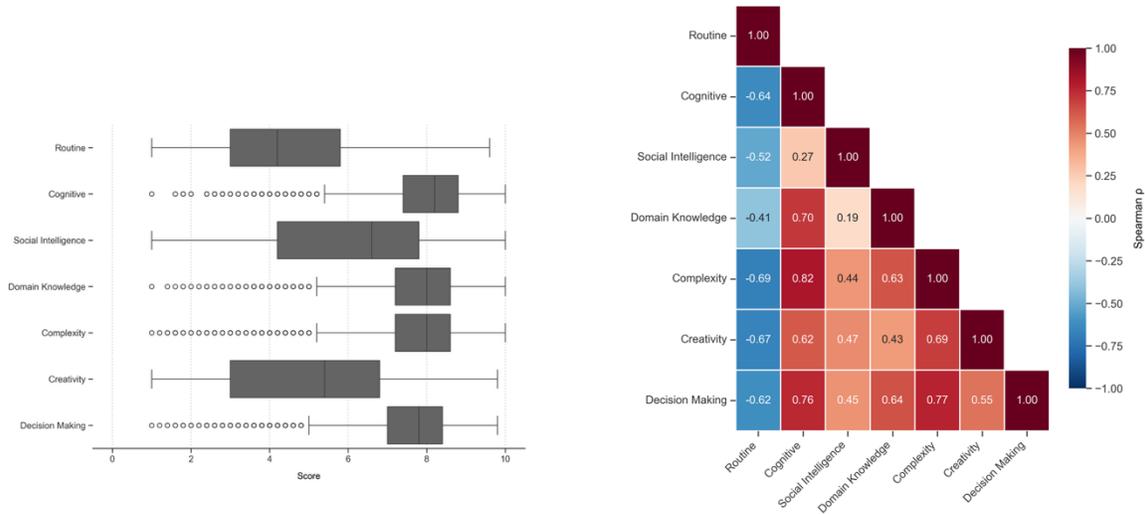

(a) Box plots showing the distribution of scores for the seven primary task characteristics.

(b) Spearman correlation heatmap illustrating the relationships between the characteristics.

Figure 2: Profile and Inter-correlation of Task Characteristics. (Source: Authors' own work)

*4.2. Which Specific Task Parameters Drive AI Adoption?*

*4.2.1. Granular Drivers of AI Usage*

To identify the most direct drivers of AI adoption, we analyzed the correlation between AI Usage and each of the 35 detailed task parameters. (Figure 3) displays the parameters most strongly associated with AI usage. The strongest positive correlations are with cr_idea_generation ($\rho = 0.173$), cg_information_processing ($\rho = 0.157$), and cr_originality ($\rho = 0.151$). Conversely, the strongest negative correlations are with rt_predictability_of_outcomes ($\rho = -0.135$) and rt_frequency_of_repetition ($\rho = -0.131$). This indicates that the share of AI usage is highest in the divergent and information-intensive phases of the work.

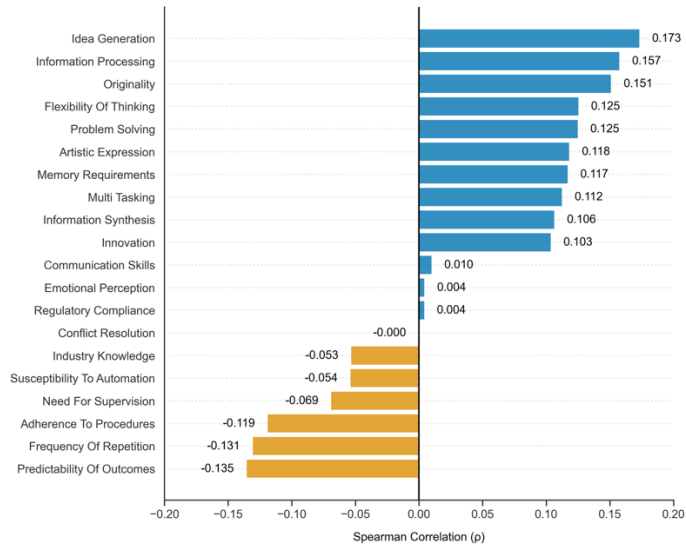

Figure 3: Spearman Correlation between Granular Task Parameters and AI Usage. Shows adoption is highest for tasks requiring idea generation and lowest for tasks characterized by procedural adherence and predictability. (Source: Authors' own work)

An analysis within the Creativity characteristic shows that the parameter for cr_idea_generation is much more strongly correlated with AI usage than cr_innovation (Figure 4b). Similarly, within the Social Intelligence characteristic (Figure 4g), all five parameters exhibit near-zero and statistically insignificant correlations with AI usage. This suggests that AI is currently used more as a tool for brainstorming than for the applied, convergent process of innovation, and also that the entire Social Intelligence domain is largely decoupled from current adoption patterns.

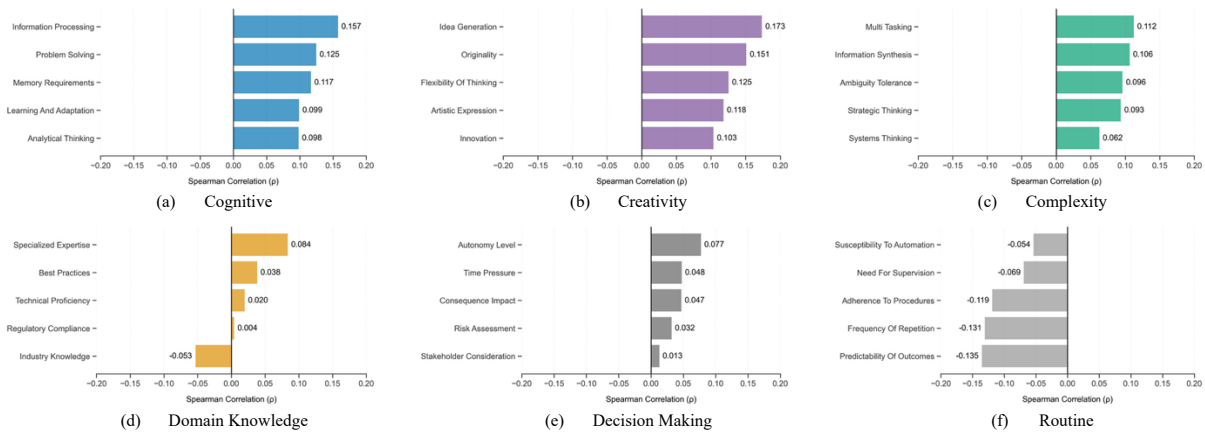

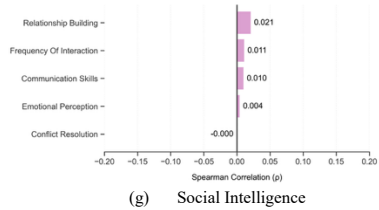

(g)   Social Intelligence

Figure 4: Intra-Characteristic Correlation Analysis. Spearman correlations of sub-parameters with AI Usage, grouped by the seven primary characteristics. (Source: Authors' own work)

### 4.2.2. Aggregate Signatures of High-Usage Tasks

When aggregated, a clear "signature" of high-usage tasks emerges. A profile comparison (Figure 5) of high-usage tasks (top 10th percentile) versus low-usage tasks (bottom 10th percentile) shows high-usage tasks score higher on Cognitive, Complexity, Creativity, and Decision Making dimensions, and significantly lower on Routine. Echoing the granular parameter analysis, the average Social Intelligence did not differ significantly between the high-usage and low-usage groups.

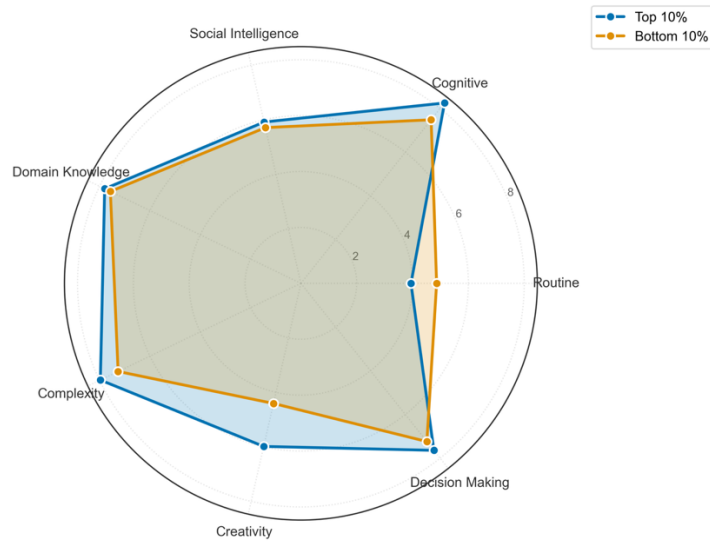

Figure 5: Mean Characteristic Scores for Tasks in the Top and Bottom Deciles of AI Usage. (Source: Authors' own work)

### 4.3. Task Archetypes and Underlying Patterns

To understand how these characteristics bundle together, we used multivariate analysis to identify task archetypes.

### 4.3.1. Latent Dimensions and Archetype Identification

We performed a Principal Component Analysis (PCA) to identify the underlying dimensions. We find the first two components explaining 82.3% of the variance. We then used K-Means clustering on these components to identify three distinct task archetypes. The choice of K=3 was supported by both the Elbow method and a superior Silhouette score (0.458) relative to solutions with more clusters. A subsequent MANOVA test confirmed that the three identified clusters were highly statistically distinct from each other across the seven characteristics (Wilks' $\Lambda = 0.112$, $p < .001$).

*4.3.2. The Three Task Archetypes*

Based on their distinct mean characteristic profiles (Table 3) (Figure 6a), we define the three archetypes as follows:

1. **Archetype 1: "Procedural & Analytical Work" (1,017 tasks):** Tasks with moderate routineness, low social intelligence, and low creativity, typical of structured analytical work.
2. **Archetype 2: "Dynamic Problem Solving" (2,100 tasks):** The largest group, defined by low routineness and the highest scores on all cognitive, creative, complexity, and decision making dimensions.
3. **Archetype 3: "Standardized Operational Tasks" (397 tasks):** A smaller group defined by extremely high routineness and the lowest scores on all other cognitive and creative dimensions.

|  | Procedural & Analytical Work | Dynamic Problem Solving | Standardized Operational Tasks |
|---|---|---|---|
| **Routine** | 5.8 | 3.36 | 7.08 |
| **Cognitive** | 7.52 | 8.53 | 4.6 |
| **Social Intelligence** | 4.51 | 7.13 | 3.1 |
| **Domain Knowledge** | 7.62 | 8.2 | 5.3 |
| **Complexity** | 7.16 | 8.43 | 4.29 |
| **Creativity** | 3.39 | 6.43 | 1.53 |
| **Decision Making** | 7.08 | 8.25 | 4.12 |

Table 3: Profile of Task Archetypes. (Source: Authors' own work)

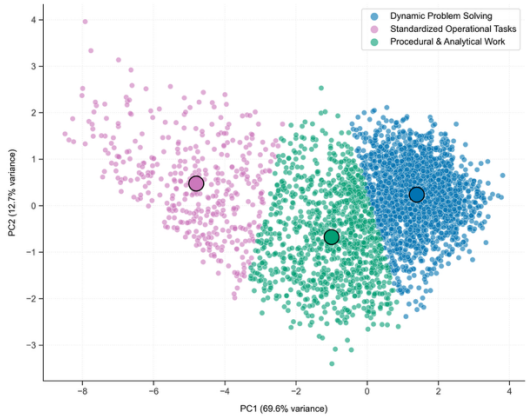 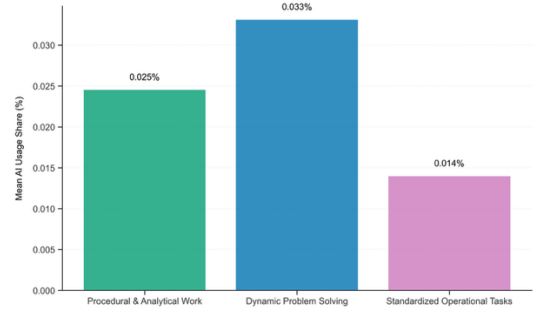

(a) Task Archetypes identified via clustering.   (b) Mean AI usage by archetype.

Figure 6: Task Archetypes. (Source: Authors' own work)

*4.4.Differential AI Adoption Across Task Archetypes*

AI adoption varies dramatically across the three task archetypes (Figure 6b). Archetype 2 ("Dynamic Problem Solving") exhibits the highest mean share of usage (3.31%). This is followed by Archetype 1 ("Procedural & Analytical Work") at 2.45%. Archetype 3 ("Standardized Operational Tasks") shows the lowest average share of usage.

An examination of the usage distribution within each cluster reveals that this pattern is driven by high-usage outliers. For all three archetypes, the distribution is extremely right-skewed; the median usage (representing the typical task) is substantially lower than the mean. The key distinction is the nature of the upper tail: the "Dynamic Problem Solving" archetype not only contains more high-usage outliers, but their magnitude is also greater than in other archetypes. Therefore, the higher average for this cluster does not reflect a broad increase in AI attention for all complex tasks, but rather a concentrated, "spiky" application of AI to a select number of them.

## 5. Discussion

Our analysis of real-world AI usage provides a nuanced portrait of the emerging human-AI division of labor. We find that AI adoption follows an extreme "long tail" distribution (with the top 5% of tasks accounting for 59% of usage), which suggests that its deployment is currently highly concentrated rather than a uniform broad-based application. This section interprets these empirical findings and explores their broader significance in understanding how generative AI is being integrated into knowledge work.

The central finding of our research suggests a significant trend towards **cognitive offloading**, where individuals leverage AI to overcome the initial high-friction stages of knowledge work, such as brainstorming, outlining, and synthesizing information. Multiple lines of evidence support this conclusion. First, tasks that attract the most AI interaction possess a distinct signature: they are defined by high levels of creativity, cognitive demand, and complexity, and a low degree of routineness. This finding aligns with recent economic research demonstrating that AI technologies are "positively correlated with cognitive analytical and interpersonal skills while negatively correlated with routine manual skills" (Colombo et al., 2024) and that "analytical non-routine tasks are at risk to be impacted by AI" (Ozgul et al., 2024). Second, granular analysis of task characteristic parameters reveals this preference at its most basic level, with AI usage correlated most strongly with task parameters such as idea generation and information processing.

Our identification of three distinct task archetypes provides a systematic framework for understanding this new division of labor. The archetype approach reveals why certain patterns emerge: it is the combination of task characteristics, rather than any single dimension, that best predicts AI's applicability. The significantly higher share of usage within the "Dynamic Problem Solving" archetype confirms that AI's primary value is currently realized in partnership with human expertise on the most demanding tasks. Conversely, the low usage in the "Standardized Operational Tasks" archetype may indicate several possibilities: that these tasks are already addressed by other forms of automation, that some highly repetitive tasks are better suited for direct API-based automation rather than conversational LLM interaction, or that current LLMs do not yet offer a compelling return on investment for such work.

A particularly insightful finding is the unique status of social intelligence. Unlike other characteristics, a task's requirement for social skill is statistically decoupled from its share of AI usage. This does not imply that AI has no social capabilities, but rather that in the current paradigm, social intelligence is not a primary factor driving adoption. This suggests that the value of human skills in empathy, negotiation, and leadership is currently not being directly amplified or substituted by AI at scale, positioning social skills as a key pillar of human comparative advantage in an AI-augmented economy.

Our findings contribute to several ongoing academic discussions. Although previous economic analyses have identified which occupations and skill categories are theoretically susceptible to AI impact, our characteristics-level analysis of actual usage patterns reveals the mechanisms driving AI adoption in practice. Furthermore, our results strongly support theories that distinguish modern generative AI from earlier automation technologies that primarily targeted routine tasks, showing instead that current AI systems are being deployed most heavily in complex, creative work.

### 5.1. Implications

These findings have significant implications across multiple domains:

1. **Labor Markets and Workforce Transformation**: Traditional white collar jobs are poised for major transformation. The evidence that people are actively choosing to delegate complex cognitive and creative tasks to AI suggests a fundamental shift in the nature of knowledge work. It is moving away from information processing and analysis, and towards task delegation, decision making, and quality evaluation. This represents a paradigm shift where execution becomes increasingly commoditized while human judgment, taste, and strategic thinking become the primary sources of value creation.
2. **Education and Human Capital Development**: The future skills landscape demands

a fundamental reorientation of educational priorities. Critical thinking, task delegation, and decision-making capabilities will become the core competencies of the AI-augmented workforce. Educational curricula must evolve beyond content delivery to focus on meta-skills: social skills necessary for clear communication with AI systems and the analytical capabilities to distinguish between valuable AI contributions and potential errors or biases. Early and frequent exposure to AI tools should become standard practice. Making AI interaction a routine part of daily life will be crucial as AI becomes increasingly integrated into all aspects of work and decision making. The future of education must prioritize AI literacy, critical thinking, and problem-solving to harness AI's potential for personalized learning while ensuring a necessary balance between AI integration and nurturing core human intelligence (Luckin, 2024; Abulibdeh, 2025).

3. **Business Strategy and Product Development**: Organizations should systematically analyze jobs that require high cognitive complexity and creativity, and develop AI-powered products and services around these functions. These are the tasks that our data show people readily offload and represent a clear market signal for where AI adoption will accelerate. They should look beyond traditional productivity metrics that focused on output volume or processing speed towards metrics that capture these higher-order contributions. Given the demonstrated willingness to delegate complex cognitive work, there should be an aggressive push toward widespread implementation of AI augmentation tools across knowledge-work environments. However, this must be coupled with robust training in critical evaluation of AI outputs, ensuring that cognitive offloading does not become uncritical acceptance. Organizations should also investigate the low adoption rates for routine tasks to determine the root cause: have these tasks already been optimized by other technologies, or are there organizational or psychological barriers, such as employees justifying their roles through time-consuming routine work?

4. **Policy and Regulatory Considerations**: Policies should focus on facilitating labor market transitions by equipping workers with the skills to operate within the high-complementarity archetypes we identify. Policymakers should consider financial incentives to accelerate workforce adaptation to AI-augmented work environments. Tax rebates or exemptions on AI skills training for companies that invest in employee AI literacy and skill training could facilitate smoother labor market transitions and reduce the risk of widespread job displacement. Achieving the necessary trust for widespread commercial deployment also requires developing standardized frameworks, such as fairness scores and certification processes, to ensure the ethical requirements of AI systems (Agarwal et al., 2023; Agarwal and Agarwal, 2024).

These findings must be interpreted in light of several limitations. First, our usage data is derived from a single family of AI models (Claude AI), whose user base may not be representative of the entire workforce. Second, our novel LLM-based scoring method for task characteristics, although systematically applied, may carry inherent biases and serves as a proxy for human judgment. Third, our analysis is cross-sectional, providing a snapshot in time; it cannot capture the dynamic evolution of AI use as technology advances and adoption patterns change. Finally, O*NET task descriptions, while comprehensive, may not capture all nuances of modern work or may group distinct activities together.

## 6. Conclusion

The question of how artificial intelligence will reshape work has been dominated by speculation. Our research addresses this through a systematic analysis of real-world AI interactions. This study reveals a critical psychological dimension of modern work: a widespread willingness to delegate the initial, and often most demanding, cognitive aspects of a task to an AI system.

This paper offers three key contributions to understanding the evolving landscape of work and AI. First, we present a methodology for systematically mapping the structure of work by breaking down broad task characteristics into quantifiable parameters. Second, we leverage this framework to demonstrate that AI's influence is highly concentrated, not evenly distributed, and primarily affects tasks that demand creativity and cognitive complexity. Third, we show that this pattern of selective impact is best captured through the lens of task archetypes: "Procedural & Analytical Work", "Dynamic Problem Solving", and "Standardized Operational Tasks".

These findings carry profound implications for how we understand and prepare for the future of work. The evidence points to a fundamental transformation in knowledge work, from information processing to task delegation, decision making, and quality evaluation. For businesses and individuals, the greatest opportunities lie in leveraging AI to tackle greater complexity and enhance creativity. For policymakers and educators, our framework provides a more nuanced guide for workforce development. It shifts the focus from a generic fear of displacement to a targeted strategy of cultivating the cognitive and social skills required for effective human-AI collaboration.

Our findings suggest that the most durable human value emerges not in direct competition with AI's cognitive capabilities, but in complementary domains. While our data demonstrates a widespread willingness to delegate complex cognitive tasks to AI systems, human advantage appears to focus on the oversight and contextual application of these AI outputs. Notably, social intelligence maintains a distinct position in this landscape, remaining statistically decoupled from AI adoption patterns and suggesting that interpersonal capabilities continue to represent an area of human comparative advantage. The implication of the cognitive offloading phenomenon is that the individuals most likely to benefit are those who can effectively navigate the partnership between human judgment and AI capability, applying their expertise to direct, refine, and contextualize AI's output rather than attempting to replicate its cognitive processing power.

**Future Research Directions**: Future research should pursue several key directions to deepen our understanding of AI-work integration. First, validation through comparable usage data from other major AI labs and longitudinal studies tracking the evolution of adoption patterns over time. Second, investigation of the psychological and organizational mechanisms driving the observed preference for delegating complex cognitive work. Third, comparison of actual AI adoption patterns with measured AI capabilities across task characteristics to identify potential misalignment between usage and optimal deployment. Finally, this research should directly inform policy frameworks for workforce development and guide AI laboratories in aligning research priorities with demonstrated real-world usage patterns.

# References

Abulibdeh, A., 2025. A systematic and bibliometric review of artificial intelligence in sustainable education: Current trends and future research directions. Sustainable Futures 10, 101033. URL: https://www.sciencedirect.com/science/article/pii/S2666188825005970, doi:https://doi.org/10.1016/j.sftr.2025.101033.

Acemoglu, D., Autor, D., 2011. Skills, tasks and technologies: Implications for employment and earnings. Handbook of Labor Economics 4, 1043–1171.

Acemoglu, D., Restrepo, P., 2019. Automation and new tasks: How technology displaces and reinstates labor. Journal of Economic Perspectives 33, 3–30.

Agarwal, A., Agarwal, H., 2024. A seven-layer model with checklists for standardising fairness assessment throughout the ai lifecycle. AI and Ethics 4, 299–314.

Agarwal, A., Agarwal, H., Agarwal, N., 2023. Fairness score and process standardization: framework for fairness certification in artificial intelligence systems. AI and Ethics 3, 267–279.

Autor, D.H., 2015. Why are there still so many jobs? the history and future of workplace automation. Journal of economic perspectives 29, 3–30.

Autor, D.H., Levy, F., Murnane, R.J., 2003. The skill content of recent technological change: An empirical exploration. Quarterly Journal of Economics 118, 1279–1333.

Barenkamp, M., et al., 2020. Applications of artificial intelligence in routine task automation. Journal of Business Research 110, 1–10.

Bessen, J.E., 2019. Automation and jobs: When technology boosts employment. Economic Policy 34, 589–626.

Bick, A., Blandin, A., Deming, D.J., 2024. The rapid adoption of generative ai. Technical Report. National Bureau of Economic Research.

Bonney, K., Breaux, C., Buffington, C., Dinlersoz, E., Foster, L., Goldschlag, N., Haltiwanger, J., Kroff, Z., Savage, K., 2024. The impact of ai on the workforce: Tasks versus jobs? Economics Letters 244, 111971. URL: https://www.sciencedirect.com/science/article/pii/S0165176524004555, doi:https://doi.org/10.1016/j.econlet.2024.111971.

Bresnahan, T.F., Trajtenberg, M., 1995. General purpose technologies 'engines of growth'? Journal of econometrics 65, 83–108.

Brynjolfsson, E., Li, D., Raymond, L., 2025. Generative ai at work. The Quarterly Journal of Economics, qjae044.

Brynjolfsson, E., Rock, D., Syverson, C., 2019. Artificial intelligence and the modern productivity paradox: A clash of expectations and statistics. The Economics of Artificial Intelligence, 23–57.

Card, D., DiNardo, J.E., 2002. Skill-biased technological change and rising wage inequality: Some problems and puzzles. Journal of Labor Economics 20, 733–783.

Colombo, E., Mercorio, F., Mezzanzanica, M., Serino, A., 2024. Towards the terminator economy: Assessing job exposure to ai through llms. arXiv preprint arXiv:2407.19204 .

Company, B.., 2024. Survey: Generative ai's up- take is unprecedented despite roadblocks. URL: https://www.bain.com/insights/survey-generative-ai-uptake-is-unprecedented-despite-bain Generative AI Survey, July 2024.

Corvello, V., 2025. Generative ai and the future of innovation management: A human centered perspective and an agenda for future research. Journal of Open Innovation: Technology, Market, and Complexity 11, 100456.

Deming, D.J., 2017. The growing importance of social skills in the labor market. Quarterly Journal of Economics 132, 1593–1640.

Deranty, J.P., 2024. Artificial intelligence and the future of work: A critical assessment. Philosophy & Technology 37, 1–25.

Eloundou, T., Manning, S., Mishkin, P., Rock, D., 2024. Gpts are gpts: Labor market impact potential of llms. Science 384, 1306–1308.

Felten, E., Raj, M., Seamans, R., 2023. How will language modelers like chatgpt affect occupations and industries? URL: https://arxiv.org/abs/2303.01157, arXiv:2303.01157.

Felten, E.W., Raj, M., Seamans, R., 2021. Occupational exposure to artificial intelligence: A new dataset and its potential uses. Strategic Management Journal 42, 2195–2217.

Fenoaltea, S., et al., 2024. Follow the money: Ai startup activity as a proxy for adoption. Journal of Economic Perspectives 38, 1–20.

Frey, C.B., Osborne, M.A., 2017. The future of employment: How susceptible are jobs to computerisation? Technological Forecasting and Social Change 114, 254–280.

Green, A., et al., 2025. Characteristics of ai job postings: Evidence from online labor markets. Labour Economics 82, 102345.

Gu, J., Jiang, X., Shi, Z., Tan, H., Zhai, X., Xu, C., Li, W., Shen, Y., Ma, S., Liu, H., et al., 2024. A survey on llm-as-a-judge. arXiv preprint arXiv:2411.15594.

Hampole, M., et al., 2025. Artificial intelligence and the demand for skills: Evidence from job postings. Review of Economics and Statistics 107, 1–15.

Handa, K., Tamkin, A., McCain, M., Huang, S., Durmus, E., Heck, S., Mueller, J., Hong, J., Ritchie,

S., Belonax, T., Troy, K.K., Amodei, D., Kaplan, J., Clark, J., Ganguli, D., 2025. Which economic tasks are performed with ai? evidence from millions of claude conversations. URL: https://arxiv.org/abs/2503.04761, arXiv:2503.04761.

Handel, M.J., 2016. Specialization and the skill premium in the 20th century. American Economic Review 106, 396–400.

Hermann, E., Puntoni, S., Morewedge, C.K., 2025. Genai and the psychology of work. Trends in Cognitive Sciences.

Idowu, A., et al., 2024. Balancing human creativity and ai assistance in creative tasks. Creativity and Innovation Management 33, 150–165.

Johnson, M., et al., 2022. Integrating ai with domain expertise: Challenges and solutions. Expert Systems with Applications 200, 116987.

Li, H., Dong, Q., Chen, J., Su, H., Zhou, Y., Ai, Q., Ye, Z., Liu, Y., 2024. Llms-as-judges: a comprehensive survey on llm-based evaluation methods. arXiv preprint arXiv:2412.05579 .

Luckin, R., 2024. Nurturing human intelligence in the age of ai: rethinking education for the future. Development and Learning in Organizations: An International Journal 39, 1–4. URL: https://doi.org/10.1108/DLO-04-2024-0108, doi:10.1108/DLO-04-2024-0108, arXiv: https://www.emerald.com/dlo/article-pdf/39/1/1/9677528/dlo-04-20

Ma, J., Wang, P., Li, B., Wang, T., Pang, X.S., Wang, D., 2025. Exploring user adoption of chatgpt: A technology acceptance model perspective. International Journal of Human–Computer Interaction 41, 1431–1445.

Ma, S., Chu, Q., Mao, J., Jiang, X., Duan, H., Chen, C., 2024. Leveraging large language models for relevance judgments in legal case retrieval. arXiv preprint arXiv:2403.18405 .

Macnamara, B., et al., 2024. Does ai enhance or replace human decision making? Cognitive Science 48, e13412.

Mathur, M., Reich, R., et al., 2024. Advancing social intelligence in ai systems: Challenges and opportunities. Nature Machine Intelligence 6, 1–3.

Miller, T., et al., 2024. Ai and domain knowledge: The importance of context. Artificial Intelligence 325, 104012.

Ozgul, P., Fregin, M.C., Stops, M., Janssen, S., Levels, M., 2024. High-skilled human workers in non-routine jobs are susceptible to ai automation but wage benefits differ between occupations. arXiv preprint arXiv:2404.06472.

Öztaş, Y.E., Arda, B., 2025. Re-evaluating creative labor in the age of artificial intelligence: a qualitative case study of creative workers' perspectives on technological transformation in creative industries. AI & SOCIETY 40, 4119–4130.

Pairolero, N.A., Giczy, A.V., Torres, G., Islam Erana, T., Finlayson, M.A., Toole, A.A., 2025.

The artificial intelligence patent dataset (aipd) 2023 update. The Journal of Technology Transfer, 1–24.

Peterson, N.G., Mumford, M.D., Borman, W.C., Jeanneret, P.R., Fleishman, E.A., Levin, K.Y., Campion, M.A., Mayfield, M.S., Morgeson, F.P., Pearlman, K., et al., 2001. Occupational information network (o*net): A comprehensive database for career exploration and job analysis. Personnel Psychology 54, 451–492.

Sarala, R.M., Post, C., Doh, J., Muzio, D., 2025. Advancing research on the future of work in the age of artificial intelligence (ai). Journal of Management Studies 62, 1863–1884.

Schroeder, K., Wood-Doughty, Z., 2024. Can you trust llm judgments? reliability of llm-as-a-judge. arXiv preprint arXiv:2412.12509 .

Singla, A., Sukharevsky, A., Yee, L., Chui, M., Hall, B., 2023. The state of ai in 2023: Generative ai's breakout year. URL: https://www.mckinsey.com/capabilities/quantumblack/our-insights/the-state-of-ai-mcKinsey Global Survey on AI.

Soori, A., et al., 2024. Ai and decision making: Supporting human judgment. Decision Support Systems 178, 114123.

Spencer, D.A., 2025. Ai, automation and the lightening of work. AI & society 40, 1237–1247.

Sweller, J., van Merriënboer, J.J., Paas, F.G., 1998. Cognitive architecture and instructional design. Educational Psychology Review 10, 251–296.

Tamkin, A., McCain, M., Handa, K., Durmus, E., Lovitt, L., Rathi, A., Huang, S., Mountfield, A., Hong, J., Ritchie, S., et al., 2024. Clio: Privacy-preserving insights into real-world ai use. arXiv preprint arXiv:2412.13678.

Tolan, S., Pesole, A., Martínez-Plumed, F., Fernández-Macías, E., Hernández-Orallo, J., Gómez, E., 2021. Measuring the occupational impact of ai: Tasks, cognitive abilities and ai benchmarks. PLOS ONE 16, e0256222.

Trammell, P., Korinek, A., 2023. Economic growth under transformative AI. Technical Report. National Bureau of Economic Research.

Tyson, L., et al., 2022. Automation and the future of work: Understanding the impact. Brookings Papers on Economic Activity 2022, 1–50.

Upreti, B., et al., 2024. The effect of ai on routine and non-routine tasks. Labour Economics 86, 102456.

Webb, M., 2020. The impact of artificial intelligence on the labor market. Quarterly Journal of Economics 135, 1945–1990.

Weinberg, B.A., 2022. Evidence from the field: The value of social skills in the ai era. Journal of Labor Economics 40, S1–S32.

Wingstrom, L., Kim, J., Lee, S., 2024. Redefining creativity in the age of ai: A task-based approach. Creativity Research Journal 36, 1–15.